\documentclass[a4paper,fleqn,usenatbib]{mnras}

\usepackage{newtxtext,newtxmath}

\usepackage[T1]{fontenc}
\usepackage{ae,aecompl}

\usepackage{graphicx}	
\usepackage{amsmath}	
\usepackage{amssymb}	

\usepackage{rotating}
\usepackage{hyperref}
\usepackage[all]{hypcap}
\hypersetup{
    bookmarks=true,         
    unicode=false,          
    pdftoolbar=true,        
    pdfmenubar=true,        
    pdffitwindow=false,     
    pdfstartview={FitH},    
    pdftitle={My title},    
    pdfauthor={Author},     
    pdfsubject={Subject},   
    pdfcreator={Creator},   
    pdfproducer={Producer}, 
    pdfkeywords={keyword1} {key2} {key3}, 
    pdfnewwindow=true,      
    colorlinks=true,        
    linkcolor=blue,         
    citecolor=blue,         
    filecolor=magenta,      
    urlcolor=cyan           
}

\title[Formation of Rotating Blowout Jet]{Flux Rope Breaking and Formation of a Rotating Blowout Jet}

\author[N.C. Joshi et al.]{
Navin Chandra Joshi,$^{1}$\thanks{E-mail: navin@khu.ac.kr, njoshi98@gmail.com}
Naoto Nishizuka,$^{2}$
Boris Filippov$^{3}$
Tetsuya Magara$^{1,4}$
\newauthor
and Andrey G. Tlatov$^{5}$
\\
$^{1}$School of Space Research, Kyung Hee University, Yongin, Gyeonggi-Do, 446-701, Korea\\
$^{2}$National Institute of Information and Communications Technology, 4-2-1, Nukui-Kitamachi, Koganei, Tokyo 184-8795, Japan\\
$^{3}$Pushkov Institute of Terrestrial Magnetism, Ionosphere and Radio Wave Propagation of the Russian Academy of Sciences (IZMIRAN), Troitsk, Moscow 142190, Russia\\
$^{4}$Department of Astronomy and Space Science, Kyung Hee University, Yongin, Gyeonggi-Do, 446-701, Korea\\
$^{5}$Kislovodsk Mountain Astronomical Station, Central Astronomic Observatory RAS, Saint Petersburg, Russia}

\date{Accepted XXX. Received YYY; in original form ZZZ}

\pubyear{2018}

\begin{document}
\label{firstpage}
\pagerange{\pageref{firstpage}--\pageref{lastpage}}
\maketitle

\begin{abstract}
We analyzed a small flux rope eruption converted into a helical blowout jet in a fan-spine configuration using multi--wavelength observations taken by \textit{SDO}, which occurred near the limb on 2016 January 9. In our study, first, we estimated the fan--spine magnetic configuration with the potential field calculation and found a sinistral small filament inside it. The filament along with the flux rope erupted upward and interacted with the surrounding fan-spine magnetic configuration, where the flux rope breaks in the middle section. We observed compact brightening, flare ribbons and post-flare loops underneath the erupting filament. The northern section of the flux rope reconnected with the surrounding positive polarity, while the southern section straightened. Next, we observed the untwisting motion of the southern leg, which was transformed into a rotating helical blowout jet. The sign of the helicity of the mini--filament matches the one of the rotating jet. This is consistent with the jet models presented by \cite{Adams14} and \cite{Sterling15}. We focused on the fine thread structure of the rotating jet and traced three blobs with the speed of 60--120 $\rm km~s^{-1}$, while the radial speed of the jet is $\sim$400 $\rm km~s^{-1}$. The untwisting motion of the jet accelerated plasma upward along the collimated outer spine field lines, and it finally evolved into a narrow coronal mass ejection at the height of $\sim$$9~R_{sun}$. On the basis of detailed analysis, we discussed clear evidence of the scenario of the breaking of the flux rope and the formation of the helical blowout jet in the fan-spine magnetic configuration.
\end{abstract}




\section{Introduction}
\label{}

Solar jets are transient plasma ejections from the lower solar atmosphere to the upper corona. They are observed in various spatial, temporal scales and wavelengths, e.g., spicules \citep{Sterling00,Sterling10,Adams14}, chromospheric jets \citep{Shibata07,Katsukawa07,Nishizuka08,Nishizuka11,Singh11}, H$\alpha$ surges \citep{Schmieder95,Uddin12}, ultraviolet/extreme ultraviolet (UV/EUV) jets \citep{Alexander99,Joshi17}, X-ray jets \citep{Shibata92,Cirtain07} and white light jets \citep{Wang02}. Jets are observed both in active regions \citep{Shimojo98,Schmieder13,Filippov13,Joshi15}, and in coronal holes \citep{Savcheva07,Young14,Sterling15}. At present, two different models are proposed to explain the jet formation: the standard jet model \citep{Shibata92} and the blowout jet model \cite{Moore10}. The second one is extended from the first one.

In the standard jet model, an emerging flux triggers a jet by magnetic reconnection with the ambient open magnetic field both in an active region and in a coronal hole, and compact bright flare loops are formed at the edge of the jet base This scenario is supported by several observations and numerical simulations \citep[e.g.,][]{Shibata92,Yokoyama95,Yokoyama96,Shimojo01,Moreno08,Nishizuka08,Archontis10}. Flux cancellation is sometimes discussed as an origin of magnetic reconnection between anti-parallel magnetic fields, instead of the emerging flux \citep{chifor08b,chifor08a}. In the blowout jet model proposed by \cite{Moore10}, an emerging flux triggers a jet in a magnetic configuration with a sheared core flux, leading to the core flux or filament eruption \cite[see also][]{Liu11,Young14}. Furthermore, \cite{Adams14} first deduced that the eruption of a mini-filament or a twisted flux rope, which is triggered by flux convergence or cancellation without a bipole emergence, is responsible for a blowout macrospicule jet in the coronal hole, and \cite{Sterling15} first deduced the same model for all or nearly all blowout jets in coronal holes. \cite{Panesar16} found that the same model fits blowout jets in non-coronal-hole regions such as quiet regions.

The blowout jets sometimes grow into small and narrow coronal mass ejections (CMEs) \citep{Hong11,Shen12,Li15}. It is also found that a helical blowout jet is originated from a twisted filament \citep{Hong13,Moore15,Filippov15}. Various numerical simulations have been performed to understand the nature of helical blowout jets \citep{Archontis10,Moreno13,Archontis13,Pariat15,Lee15}. A twisted flux rope is produced by the internal magnetic reconnection between sheared field lines in an emerging flux region \citep{Archontis13}. The flux rope is stimulated to erupt by the kink or torus instability, leading the external reconnection \citep{Moreno13}. This process produces a jet and the untwisting motion of the twisted flux tubes \citep{Li15}. \cite{Pariat15} considered another configuration of a fan-spine structure, which stored the energy of the shared field at a separatrix, and reproduced a helical jet via magnetic reconnection at the separatrix surface. The data-driven Magnetohydrodynamic (MHD) simulation also shows that the emerging flux near the fan-spine configuration can produce a helical jet \citep{Cheung15}.

Historically, a helical motion was first found in macrospicules. They were observed with an EUV spectrometer. Their properties are high temperature $\rm (2-3) \times 10^{5}~K,~30-60~Mm$ height, fast flows at the speed of 80 $\rm km~s^{-1}$, high density $\rm 10^{10}~cm^{-3}$, brighter edges than the central region, and the broad emission line which narrows in the later phase \citep{Pike97,Parenti02}. There are statistical studies of their properties \citep{Yamauchi05,Bennett15,Kiss17}. The horizontal unfolding motion of a twisted flux rope shows Doppler shifts on the limb, at the speed of 50-120 $\rm km~s^{-1}$, in association with an independent radial flow of the jet \citep{Pike98,Scullion09,Kamio10}. The helical motion is also discussed on the base of imaging observations \citep{Canfield96,Jiang07,Patsourakos08,Shen11,Lee13}, as well as the coexistence of a macrospicule and a blowout jet \citep{Kayshap13,Adams14}.

In this paper, we present a clear event of a helical blowout jet with the detailed analysis of its formation and initiation. We discuss the following questions in details: what is the origin of the rotation in blowout jets and how is the helicity transferred? In Section~\ref{sec2}, we describe the observational dataset. In Section~\ref{sec3}, we explain topologies of the filament and the surrounding magnetic field configuration. In Section~\ref{sec4}, we present our observational analysis of the helical blowout jet and the associated CME. In Section~\ref{sec5}, we summarize our results and discuss the origin of an untwisting motion of helical blowout jets.

\section{Observational Data Set}
\label{sec2}
We used observational data taken by the Atmospheric Imaging Assembly \citep[AIA;][]{Lemen12} on board {\it the Solar Dynamics Observatory} \citep[{\it SDO};][]{Pesnell12}. AIA takes full--disk images of the Sun with seven extreme ultraviolet (EUV), two ultraviolet (UV) and one continuum wavelength channels with the common spatial resolution of $0.6\arcsec$ per pixel. The temporal resolutions of UV and EUV observations are 24 and 12 s, respectively. The time-series of EUV data taken with 304 \AA\ and 171 \AA\ filters of
AIA/\textit{SDO} are used to study a mini--filament eruption and an associated jet, in addition to H$\alpha$ data taken by the National Solar Observatory-Global Oscillation Network Group (NSO-GONG). We used the line--of--sight (LOS) magnetograms taken by the Helioseismic and Magnetic Imager \citep[HMI;][]{Schou12} on board \textit{SDO}. It provides the photospheric magnetic field data with the resolution of $0.5\arcsec$ per pixel with a minimum cadence of 45 s. 

White light data is taken by the Large Angle and Spectrometric Coronagraph \citep[LASCO;][]{Burueckner95}, which consists of three coronagraphs C1, C2, and C3, on board the \textit{Solar and Heliospheric Observatory} (\textit{SOHO}). We used the white light coronagraph data of C2 and C3. 


\section{Magnetic Topology of the Filament and the Surrounding Field}
\label{sec3}


\subsection{Initial Magnetic Configuration of the H$\alpha$ mini-filament and the Ambient Coronal Field}
\label{sec3.1}

\begin{figure}
\vspace*{-4cm}
\centerline{
	\hspace*{0.5\textwidth}
	\includegraphics[width=1.1\textwidth,clip=]{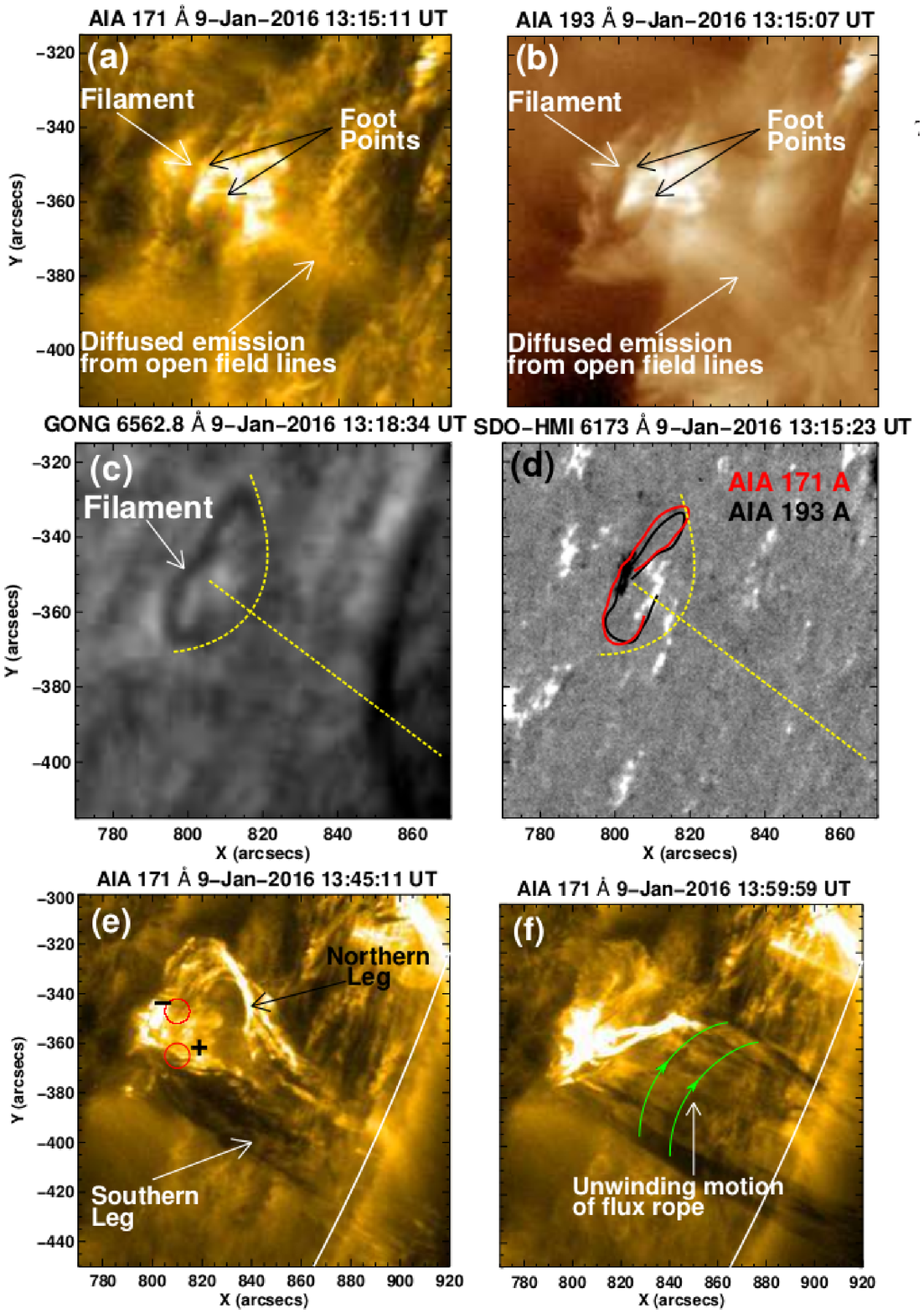}
	}
\vspace*{-2cm}
\caption{((a)--(c)) Snapshot images of a mini-filament near the limb on 2016 January 9, taken by 171 \AA\ and 193 \AA\ filters of AIA/\textit{SDO} and H$\alpha$ filter of NSO/GONG. (d) A magnetogram of the photosphere taken by HMI/\textit{SDO}. ((e)--(f)) The eruption of mini--filament and unwinding motion of the right-- handed helix was observed at 13:45:11 UT and at 13:59:59 UT, respectively. The separator lines are drawn by hand, based on the potential filed calculation (see Figure~\ref{fig3}(b)).}
\label{fig1}
\end{figure}

Figure~\ref{fig1} shows snapshot images of a mini-filament near the west limb on 2016 January 9, the magnetic field at the footpoint of the filament, and its time evolution. Figures~\ref{fig1}(a)--(b) represent a dark mini-filament taken with the AIA 171 \AA\ and 193 \AA\ filters at $\sim$13:15 UT, while Figure~\ref{fig1}(c) shows a corresponding H$\alpha$ filament taken by GONG/NSO. The mini--filament is elongated from the northwest to the southeast. The magnetic field in the same area taken by HMI/\textit{SDO} is shown in Figure~\ref{fig1}(d).

\begin{figure}
\vspace*{-6cm}
\centerline{
	\hspace*{0.6\textwidth}
	\includegraphics[width=1.2\textwidth,clip=]{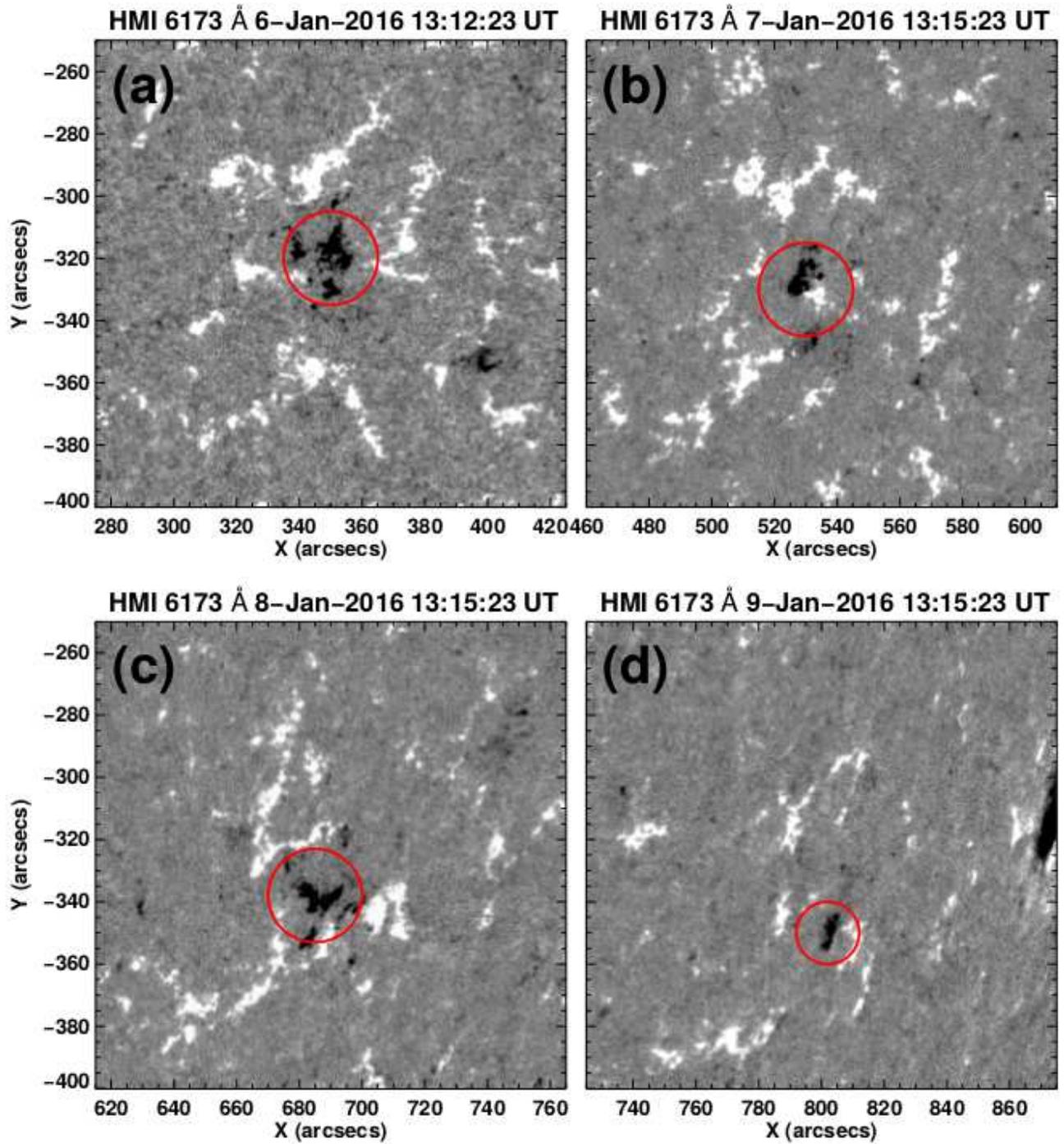}
	}
\vspace*{-5cm}
\caption{Photospheric magnetogram images were taken by HMI/\textit{SDO} during 2016 January 6--9. The central negative mono-pole region is marked by red circles.}
\label{fig2}
\end{figure}

The filament was close to the limb on January 9, where the photospheric magnetic field measurement is obscure due to the projection effect. Figure~\ref{fig2} represents snapshots of the line-of-sight magnetograms taken by HMI/\textit{SDO} during 2016 January 6--9. During the period, we found a patch of the negative polarity surrounded by the positive polarity, and the overall photospheric magnetic field structure is kept almost the same.

\begin{figure}
\vspace*{-8cm}
\centerline{
	\hspace*{0.55\textwidth}
	\includegraphics[width=1.15\textwidth,clip=]{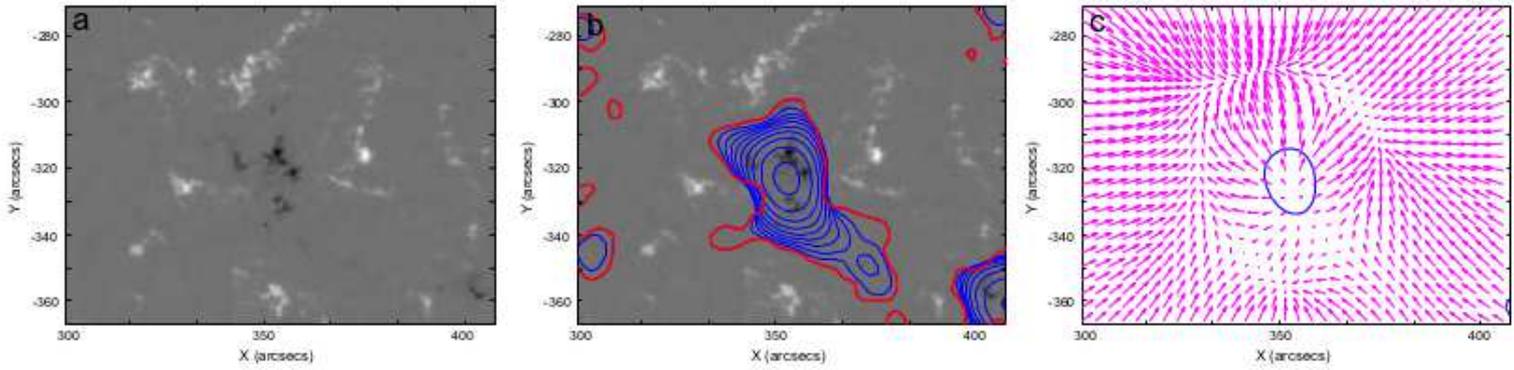}
	}
\vspace*{-11cm}
\caption{((a)--(b)) The magnetogram image were taken by HMI/\textit{SDO}, at 13:12 UT on 2016 January 6, with contours of PILs at different heights colored in red and blue. The red contour is at the height of 1.5 Mm, while blue contours are at the height of 3.0-15.0 Mm with the step of 1.5 Mm. (c) The map of the horizontal field at the height of 13.5 Mm with the PIL.}
\label{fig3}
\end{figure}

To calculate the potential magnetic field above the region, we used the magnetogram at 13:12 UT on 2016 January 6 (Figure~\ref{fig3}(a)). For potential-field calculations, we used the Green function method to solve numerically the Neuman external boundary value problem \citep[see][]{Filippov01,Filippov13}. This method is most preferable for the extrapolation of the photospheric magnetic field within a restricted region into the corona. Figure~\ref{fig3}(b) shows polarity inversion lines (PILs) at different heights. The red contours present the PIL at the height of 1.5 Mm, while the blue contours present the PIL at heights from 3 Mm to 15 Mm, with the step of 1.5 Mm. Above the height of 15 Mm, the field is upwardly directed everywhere, and a null point is located just below this height. Figure~\ref{fig3}(c) shows the map of the horizontal magnetic field at the height of 13.5 Mm with the contour of the PIL. The 3D null point is located on the PIL and at the point where the horizontal magnetic field diverges and vanishes.

\begin{figure}
\vspace*{-8cm}
\centerline{
	\hspace*{0.55\textwidth}
	\includegraphics[width=1.15\textwidth,clip=]{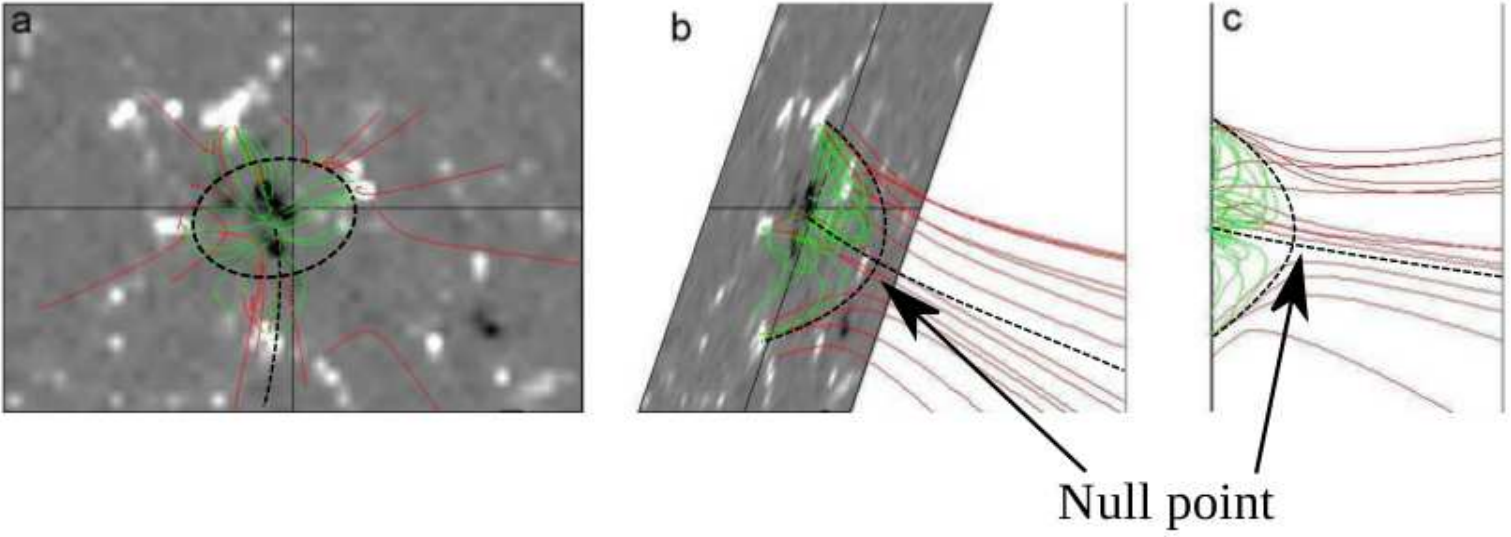}
	}
\vspace*{-11cm}
\caption{The calculated potential magnetic field lines of the corresponding active region on 2016 January 9, viewed from the top (a), from the top-side (b), and from the side (c). The open field lines are colored in red, while the closed field lines are in green. The expected separatrices and PIL are overlaid with dotted black lines.}
\label{fig4}
\end{figure}

The potential field structure is shown in Figure~\ref{fig4} with different views. All the field lines emanating from the central negative polarity are connected to the surrounding positive polarity region with closed structure. Outside of this central area, all field lines are open. In a side view, it is found that the structure is a fan-spine configuration. The top of the dome which covers closed lines is at the height of about 13-15 Mm, which is higher than the estimated height of the mini filament.

The approximate separatrix of this configuration based on the potential field calculation are overlaid in Figures~\ref{fig1}(c) and~\ref{fig1}(d) with yellow dashed lines. Along the spine field line, there seems a diffused emission in AIA/\textit{SDO} 171 \AA\ and 193 \AA\ filter images (Figure~\ref{fig1}(a) and~\ref{fig1}(b)).


\subsection{Anchoring and Chirality of the Filament}
\label{sec3.2}

We derived the filament shape from the AIA/\textit{SDO} 171 \AA\ and 193 \AA\ images (Figures~\ref{fig1}(a) and~\ref{fig1}(b)) and over--plotted them on the magnetogram colored in red for 171 \AA\ and in black for 193 \AA (see Figure~\ref{fig1}(d)). The northern footpoint of the filament is anchored in the negative polarity, while the southern footpoint is connected with the positive polarity. This anchoring is confirmed by AIA images of the mini-filament eruption at 13:45:11 UT, when the mini-filament is transformed into a jet (Figure~\ref{fig1}(e)). Since only the leg anchored in the positive polarity can merge into the ambient positive flux and produce a jet into the corona, the southern leg connected to the jet should be anchored in the positive polarity. Accordingly, the northern leg is anchored in the negative polarity.

According to this anchoring, the axial magnetic field of the mini--filament is directed from the south to the north, i.e., leftward when it is viewed from the positive polarity (from the east). 
On the basis of the direction of the filament axial magnetic field, we identified the filament chirality to be sinistral. In flux rope models, sinistral filaments lie inside right--handed helical fields \citep{Martin98,Martin03}. During the rotating phase of the jet at 14:00 UT, we actually observed the unwinding of a right--handed helix of the flux rope (shown by green curved lines in Figure~\ref{fig1}(f)) (see the section~\ref{sec4} for more details). The observed right--handed twist of the flux rope helix is consistent with its sinistral chirality. This case also seems to be in agreement with the general hemispheric chirality rule \citep{Pevtsov03}.  


\section{Observational Analysis of the Jet Formation and Dynamics}
\label{sec4}

\begin{figure}
\vspace*{-3.5cm}
\centerline{
	\hspace*{0.5\textwidth}
	\includegraphics[width=1.1\textwidth,clip=]{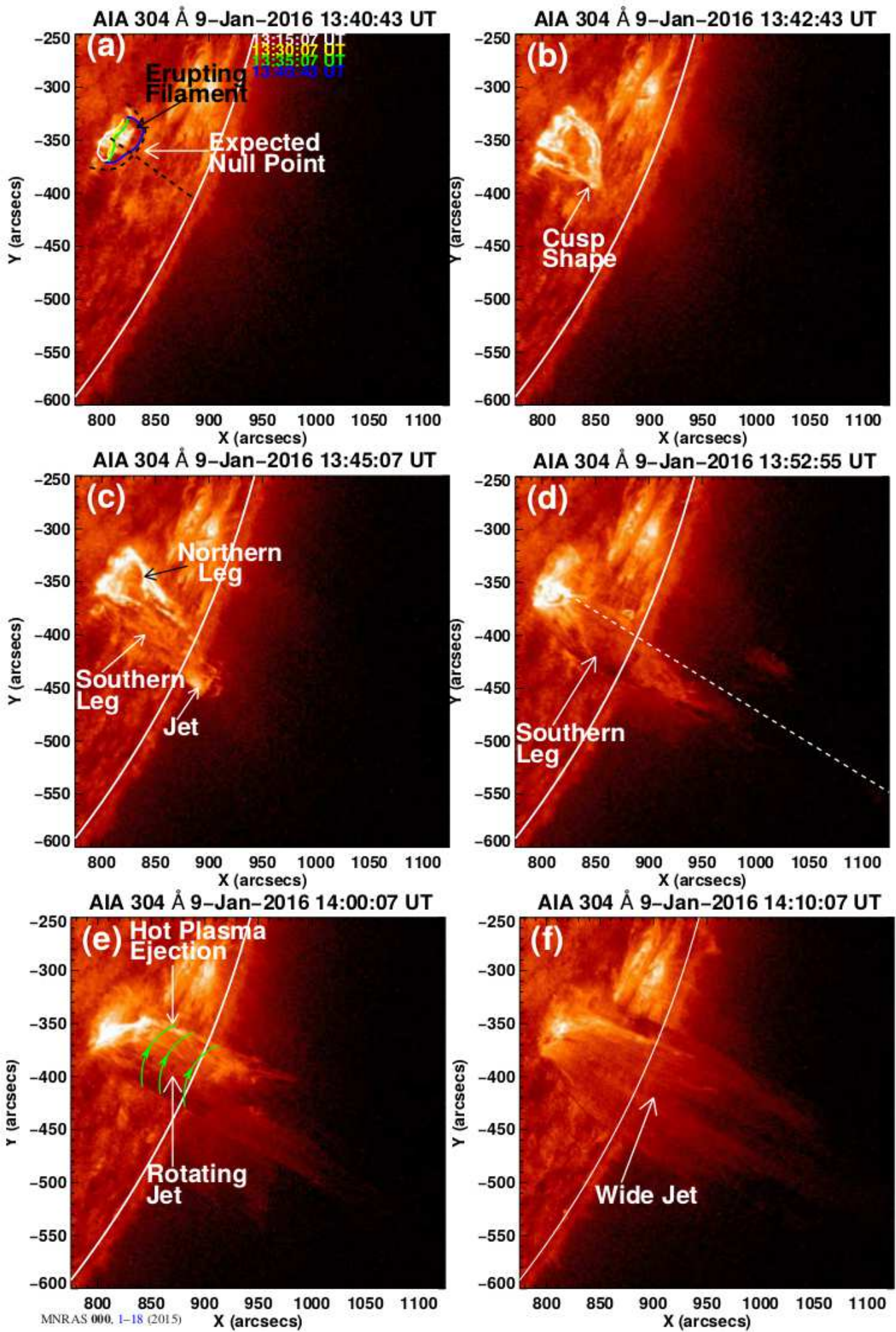}
	}
\vspace*{-0.8cm}
\caption{Snapshot images of the dynamics of the helical blowout jet formation taken by 304 \AA\ filter of AIA/\textit{SDO}. The limb is shown by a solid white line.}
\label{fig5}
\end{figure}

Figure~\ref{fig5} shows the snapshots of the helical blowout jet observed in the 304 \AA\ channel of AIA on 2016 January 9. The mini--filament was observed for many hours before 13:15 UT. The mini--filament slowly rose upward at 13:34 UT. The filament approached to the magnetic null point, keeping its arch shape (shown by different colors for each step in Figure~\ref{fig5}(a)). The fan--spine structure and the location of the magnetic null, which were estimated from the potential field calculation, are drawn in Figure~\ref{fig5}(a). The filament continued to move upward up to $\sim$13:42 UT, when the arch shape changed into a cusp--like structure (Figure~\ref{fig5}(b)).

\begin{figure}
\vspace*{-3cm}
\centerline{
	\hspace*{0.5\textwidth}
	\includegraphics[width=1.1\textwidth,clip=]{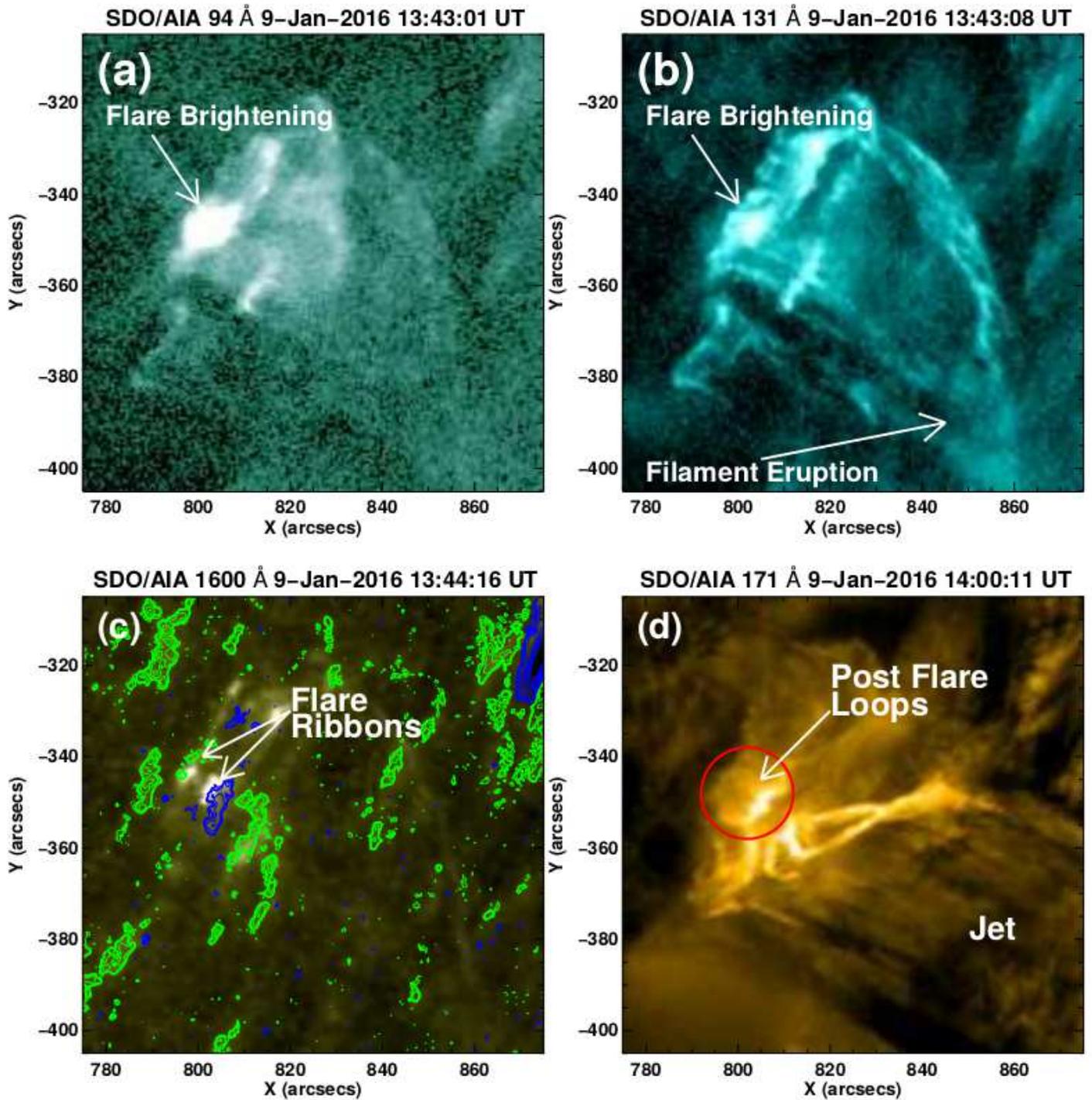}
	}
\vspace*{-4cm}
\caption{((a)--(b)) The compact brightening below the filament eruption observed by 94 \AA\ and 131 \AA\ filters of AIA/\textit{SDO} at $\sim$13:43 UT on 2016 January 9. (c) The two compact ribbons--like structures taken by 1600A \AA\ filter of AIA/\textit{SDO} at 13:44:16 UT, and (d) the post--flare loops taken by 171 \AA\ filter of AIA/\textit{SDO} at 14:00:11 UT.}
\label{fig6}
\end{figure}

Figures~\ref{fig6}(a) and~\ref{fig6}(b) show the 94 \AA\ and 131 \AA\ images of AIA/\textit{SDO} at around 13:43 UT. The 94 \AA\ and 131 \AA\ filter images represent hot plasma, and a compact brightening was observed just below the erupting filament. This compact brightening is stretched along a PIL, where the two--ribbon structure appeared at the same place in 1600 \AA\ images (Figure~\ref{fig6}(c)). One part of the two--ribbon structure was located in the positive polarity, while the other one in the negative polarity. Furthermore, in the later phase, post--flare loops were formed above the two--ribbon structure, which was observed in 171 \AA\ images (Figure~\ref{fig6}(d)). In the standard flare model, the two--ribbon  structure is produced by the bombardment of the chromosphere by released electron beams from the magnetic reconnection point in the corona just below an erupting flux rope, and the post-flare loops are formed by the chromospheric evaporation after the bombardment (see CSHKP model: \cite{Carmichael64}, \cite{Sturrock66}, \cite{Hirayama74}, \cite{Koop76}). By analogy, we expect magnetic reconnection to occur below the erupting mini-filament, which released high energy particles and caused the compact brightening and the two-ribbon structure.

After 13:45 UT the ascending arch of the erupting filament seems to break into two semi-arches: with the southern and northern parts. The southern part with the footpoint anchored in the positive polarity proceeds ascending motion and transforms into a rotating column i.e., a jet. The magnetic field of this leg is upwardly directed, which is coincided with the direction of the ambient coronal field. Therefore, the internal field of this leg can easily merge into the ambient coronal field. The northern part of the filament stopped its ascending motion and began to move downward. Since the foot point of this leg is anchored in the negative polarity, its axial field becomes anti--parallel to the coronal field at the height above the fan surface. The field of the northern leg cannot be further stretched into the corona because antiparallel field lines should reconnect. It seems that the northern part of the filament found anchoring for the other footpoint in the nearby positive polarity (Figure~\ref{fig5}(c)), forming an arch shape for a few minutes and then disappeared (see Figure~\ref{fig5}(c) and attached movie). This arch shape is longer than the post-flare loops above the mini two-ribbon structure.

\begin{figure}
\vspace*{-4cm}
\centerline{
	\hspace*{0.6\textwidth}
	\includegraphics[width=1.2\textwidth,clip=]{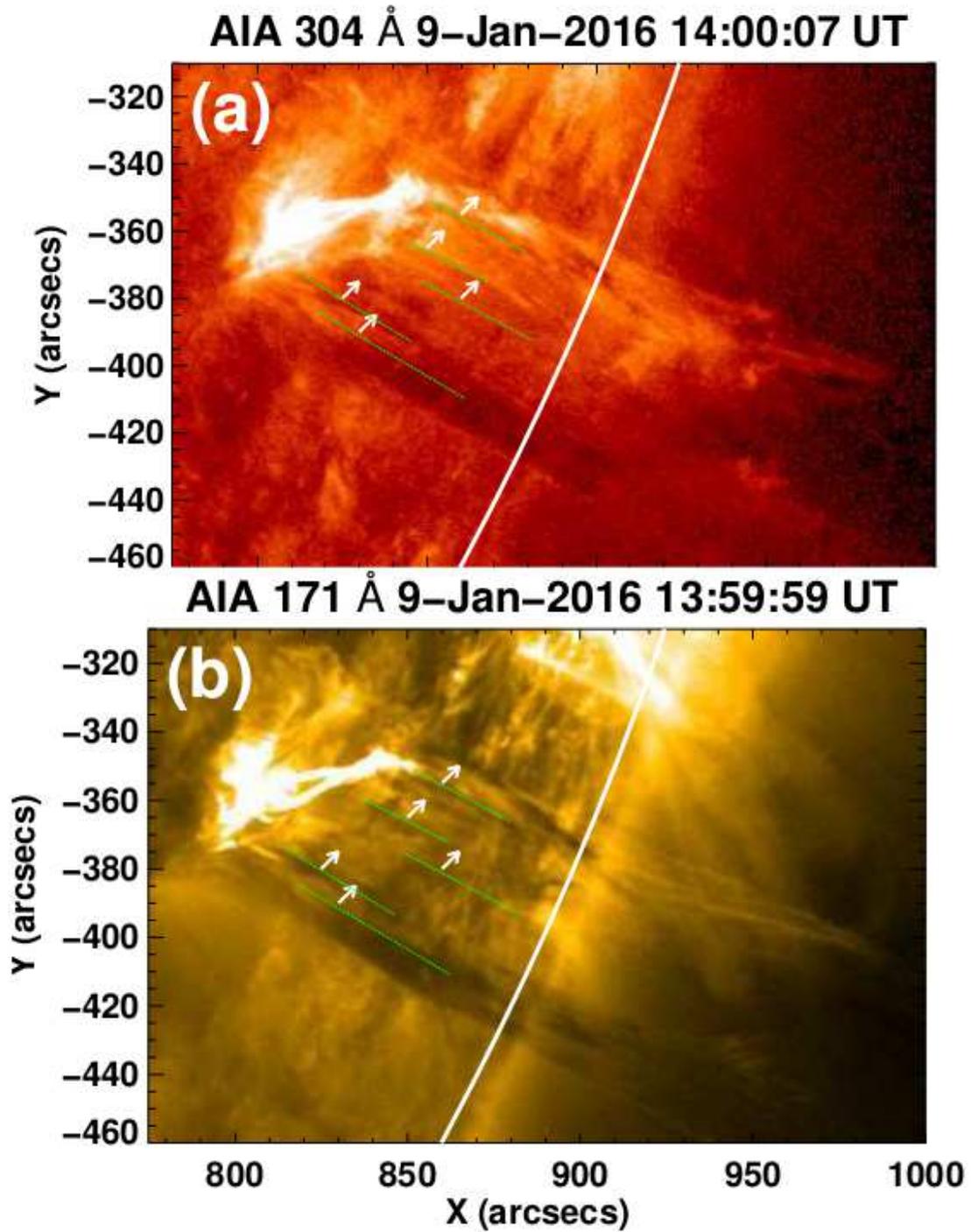}
	}
\vspace*{-5.5cm}
\caption{((a)--(b)) The fine threads of the rotating jet observed by 304 \AA\ and 171 \AA\ filters of AIA/\textit{SDO} at $\sim$14:00 UT on 2016 January 9, marked by dotted green lines. The directions of the motion are marked by small white arrows in the panels.}
\label{fig7}
\end{figure}

At 13:53 UT, the southern part of the filament stretched along the open collimated path is visible (Figure~\ref{fig5}(d)). The rotational motion started at around 13:52 UT and ended at around 14:07 UT. The direction of the rotation is from the southwest to the northeast (when viewed on disk), and in other words, a clockwise rotation when viewed from the above. The direction of the rotation motion is represented by green arrows in Figure~\ref{fig5}(e). Figure~\ref{fig7} shows zoom-up images of the fine structure of the rotating jet, observed with AIA 304 \AA\ and 171 \AA\ filters at $\sim$14:00 UT. We identified some fine dark threads moving as a whole from the southeast to the northwest on the front side of the jet. Five clear fine threads are marked by dotted green lines with their directions of motion by white arrows. At last, most of the threads became stretched radial (see also Figure~\ref{fig5}(f)).

We also found a cusp-like hot plasma ejection from 13:53 UT to 13:58 UT, which lasted for a few minutes (Figure~\ref{fig5}(e)). This is similar to a plasmoid ejections
in solar flares \citep{Shibata94} and tiny plasmoid ejections in chromospheric jets \citep{Singh12}. Another possibility is a corresponding structure of a propagating shock driven by magnetic reconnection and the associated velocity pulse, as proposed in spicules and macrospicules \citep[e.g.,][]{Shibata82,Kayshap13}.

\begin{figure}
\vspace*{-6cm}
\centerline{
	\hspace*{0.6\textwidth}
	\includegraphics[width=1.2\textwidth,clip=]{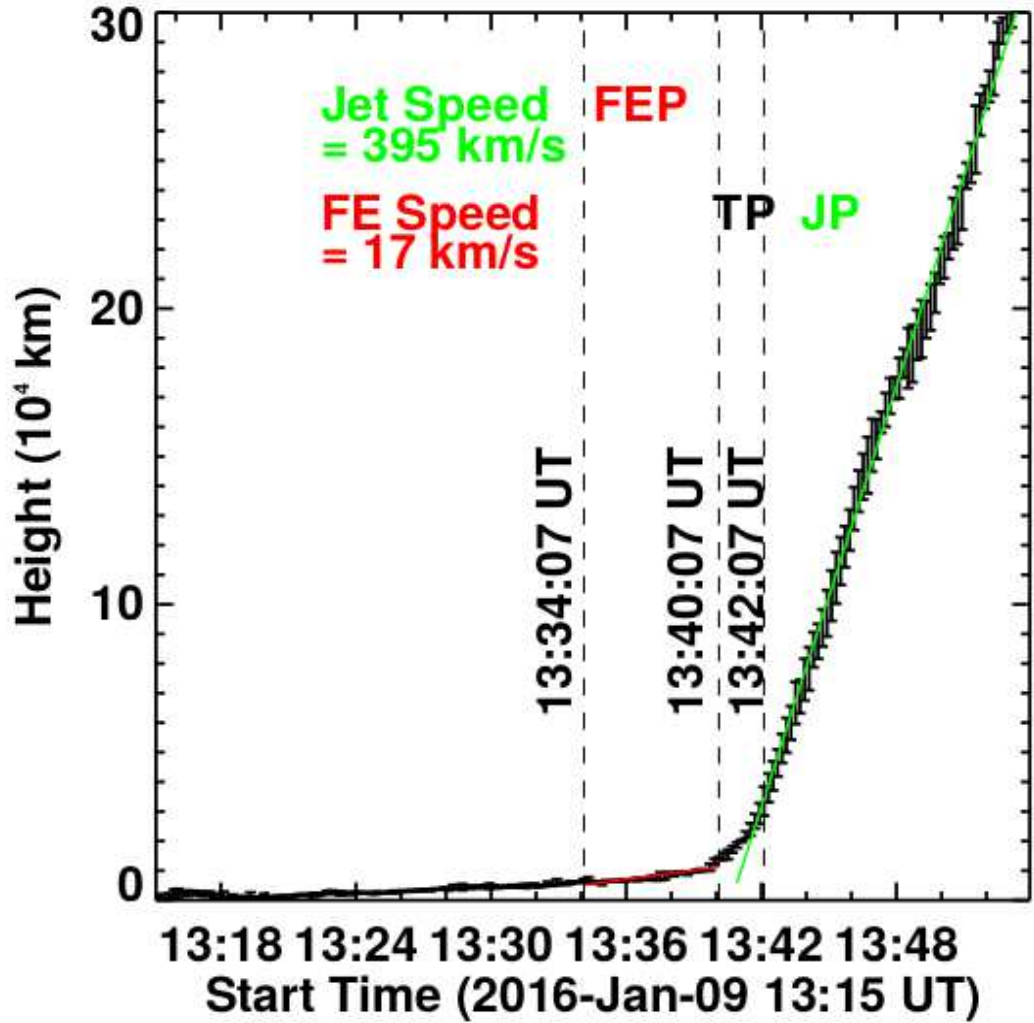}	
	}
\vspace*{-8cm}
\caption{The height--time profile of the filament eruption transformed into a jet from 13:15 UT to 13:53 UT. The attached error bars are calculated by the standard deviations of the three times measurements. The linear fitting shows the speed of the filament eruption 17$\pm2$ $\rm km~s^{-1}$ and the jet velocity 395$\pm3$ $\rm km~s^{-1}$, respectively.}
\label{fig8}
\end{figure}

We measured the upward velocity of the filament and the jet from a height--time diagram (Figure~\ref{fig8}). The slit is shown in Figure~\ref{fig5}(d). We tracked the front edge of the erupting filament and the jet. The speed was calculated by the linear fitting. The profile shows the initial slow eruption of the filament with an average speed of 17$\pm 2$ $\rm km~s^{-1}$, during 13:34 UT--13:40 UT, and the acceleration to a higher speed of 395$\pm 3$ $\rm km~s^{-1}$ during 13:42 UT--13:53 UT. The acceleration is synchronized with the transformation of the filament into the jet between 13:40 UT and 13:42 UT.

\begin{figure}
\vspace*{-3cm}
\centerline{
	\hspace*{0.5\textwidth}
	\includegraphics[width=1.1\textwidth,clip=]{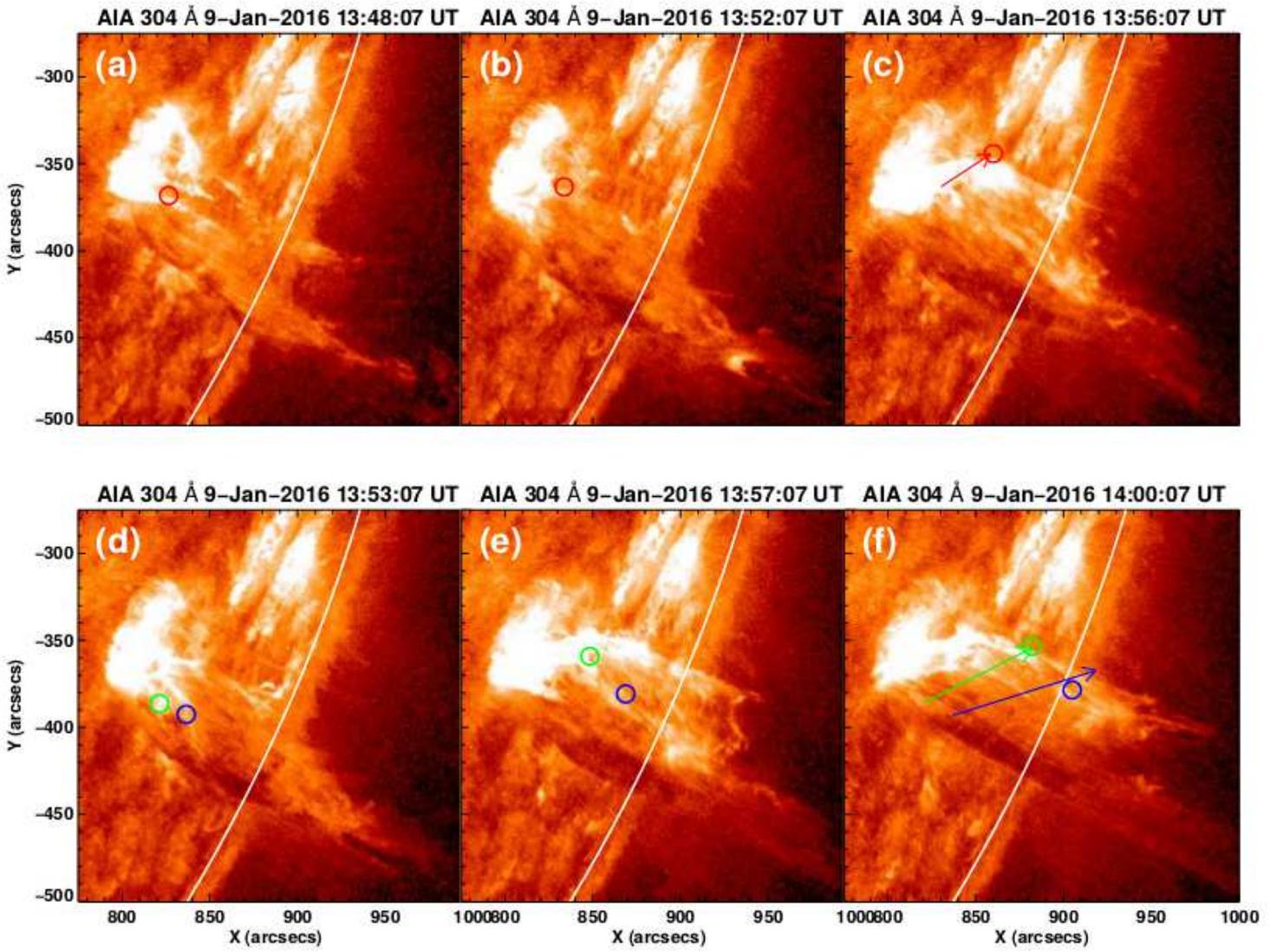}
	}
\vspace*{-6cm}
\caption{Snapshot images of the rotating jet with three moving plasma blobs colored in red, blue and green, observed by 304 \AA\ filter of AIA/\textit{SDO}.}
\label{fig9}
\end{figure}

\begin{figure}
\vspace*{-6cm}
\centerline{
	\hspace*{0.6\textwidth}
	\includegraphics[width=1.2\textwidth,clip=]{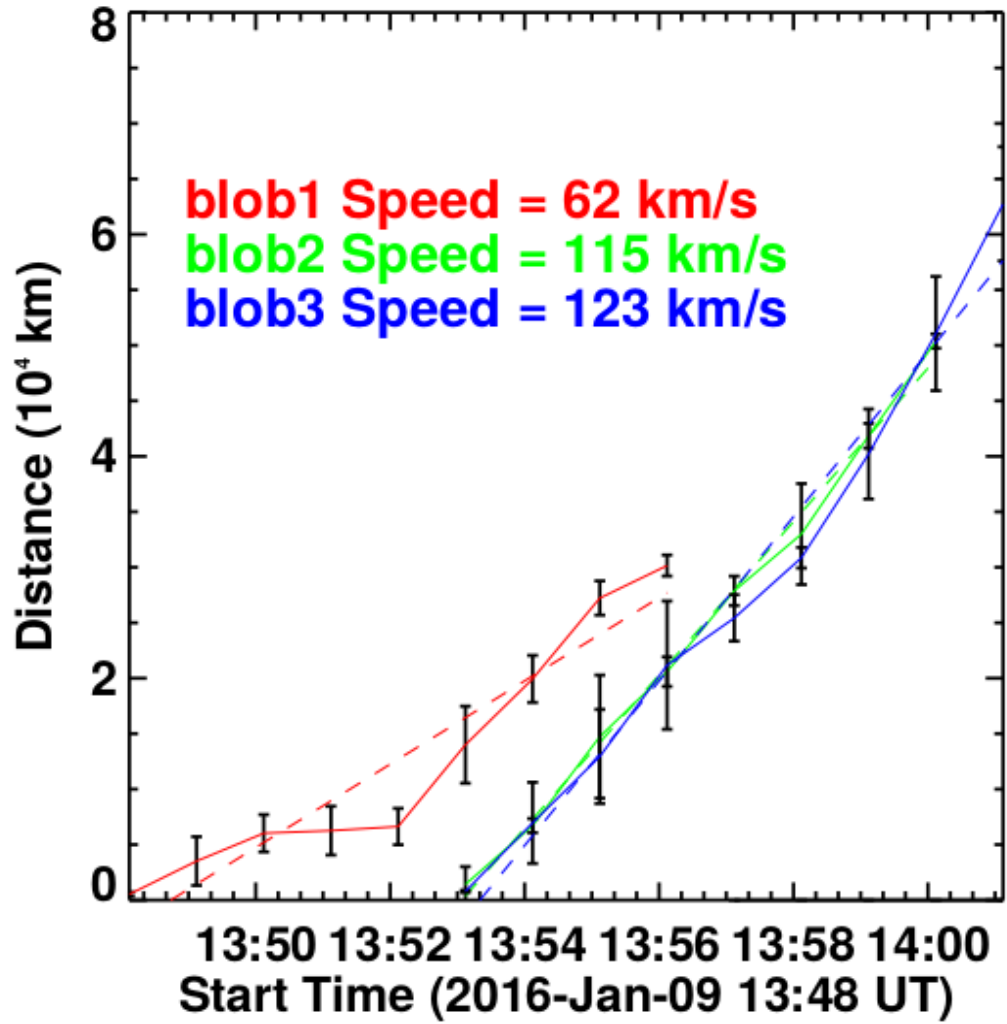}	
	}
\vspace*{-8cm}
\caption{The distance--time plots of the rotating bright plasma blobs along the trajectories shown in Figures~\ref{fig9}(c) and~\ref{fig9}(f). The linear fitting shows the speed of rotation motions of three blobs, 62$\pm$7 $\rm km~s^{-1}$, 115$\pm$3 $\rm km~s^{-1}$ and 123$\pm$6 $\rm km~s^{-1}$.}
\label{fig10}
\end{figure}

The rotation motion of the jet is analyzed in more detail in Figures~\ref{fig9} and~\ref{fig10}. Figure~\ref{fig9} shows the images of AIA 304 \AA\ from 13:48 UT to 14:00 UT. We tracked three small plasma blobs (or bright spots) moving in front of the jet axis from the southeast to the northwest. These three moving bright blobs are shown by the red, green and blue circles. The coordinates of the starting points are used as the reference points. Then we measured the distance between a reference point and a rotating blob at different times. To estimate the errors in the measurements, we repeated measurements three times for each blob and calculated the standard deviation of the three distances for an error. The distance-time diagrams of the three rotating blobs are shown in Figure~\ref{fig10}. The linear fitting method was used to estimate the blob velocity. It is found that the different parts of the jet rotate with an average speed of 62$\pm$7 $\rm km~s^{-1}$, 115$\pm$3 $\rm km~s^{-1}$, and 123$\pm$6 $\rm km~s^{-1}$. The estimated speeds include both horizontal and vertical motions because the blobs move both in the vertical and horizontal directions. These rotation speeds are much lower than the jet eruption speed.

\begin{figure}
\vspace*{-5cm}
\centerline{
	\hspace*{0.54\textwidth}
	\includegraphics[width=1.15\textwidth,clip=]{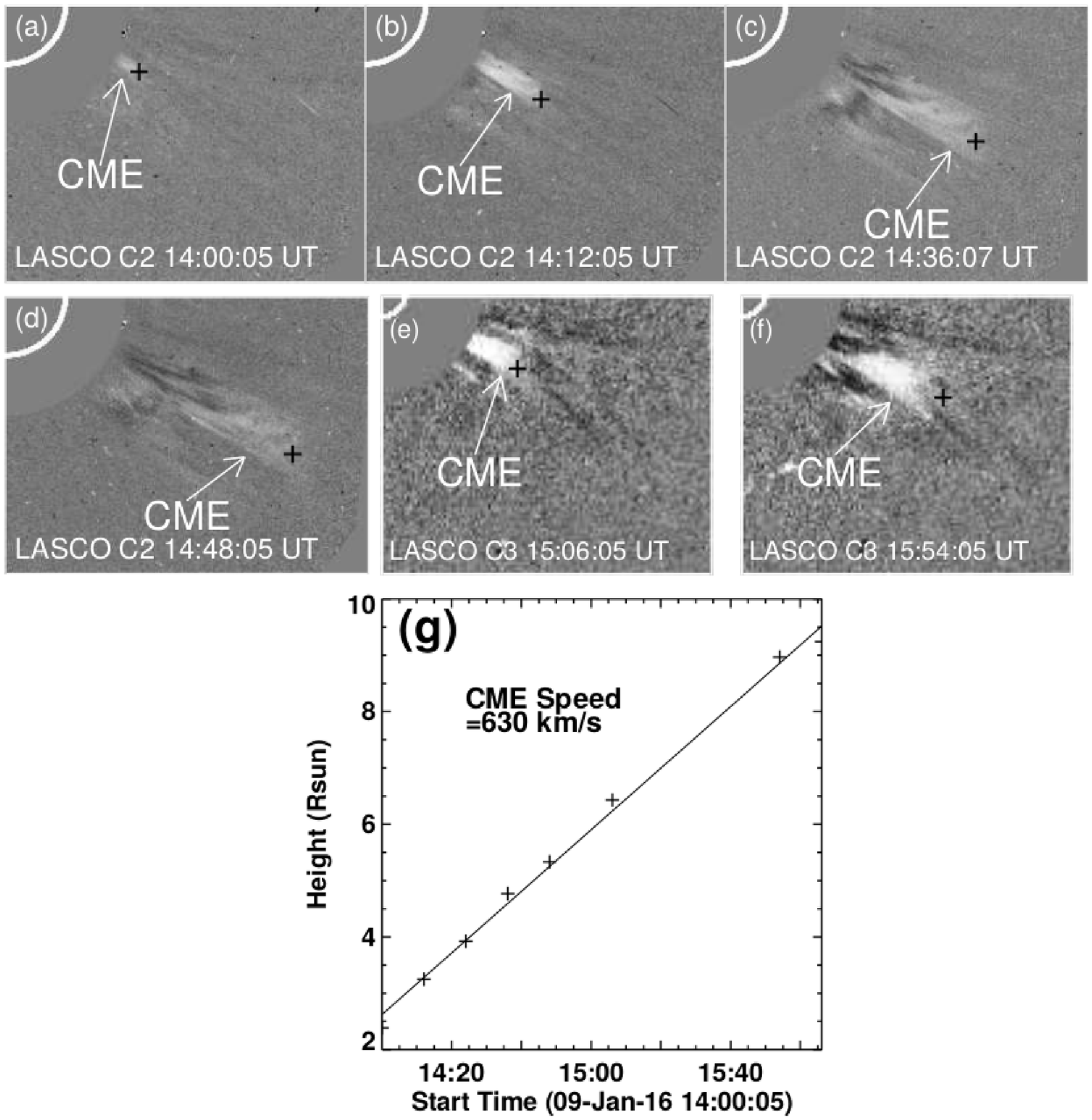}
	}
\vspace*{-4cm}
\caption{((a)-(f)) The coronagraph images of the propagation of the narrow CME associated with the helical blowout jet, taken by the C2 and C3 telescopes of LASCO on board \textit{SOHO}. The images were downloaded from the site of LASCO/\textit{SOHO} catalog (\url{http://cdaw.gsfc.nasa.gov/CME\_list/}). (g) The height--time profile of the narrow CME. The linear fitting shows the speed of 630 $\rm km~s^{-1}$ with an error of $\pm$18 $\rm km~s^{-1}$.}
\label{fig11}
\end{figure}

The jet was associated with a narrow CME (Figure~\ref{fig11}). The CME was observed after 14:00 UT by the LASCO C2 coronagraph. The CME was narrow and collimated (Figure~\ref{fig11}(b)). The width of the CME slightly increased with time (Figure~\ref{fig11}(c)--~\ref{fig11}(f)). The CME appeared in the LASCO C3 field of view at 15:06 UT (Figure~\ref{fig11}(e)) and continued to move upward (Figure~\ref{fig11}(f)). Figure~\ref{fig11}(g) shows the height--time profile of the leading edge of the CME from 14:00 UT to 16:00 UT. To estimate the speed, we fitted the data points with a straight line. The speed of the narrow CME was $\sim$630 $km~s^{-1}$. It indicates the process of acceleration of the jet material from the speed of $\sim$400 $km~s^{-1}$ to $\sim$600 $km~s^{-1}$.


\section{Results and Discussion} 
\label{sec5}

\begin{figure}
\vspace*{-4.5cm}
\centerline{
	\hspace*{-0.5\textwidth}
	\includegraphics[width=1.2\textwidth,clip=]{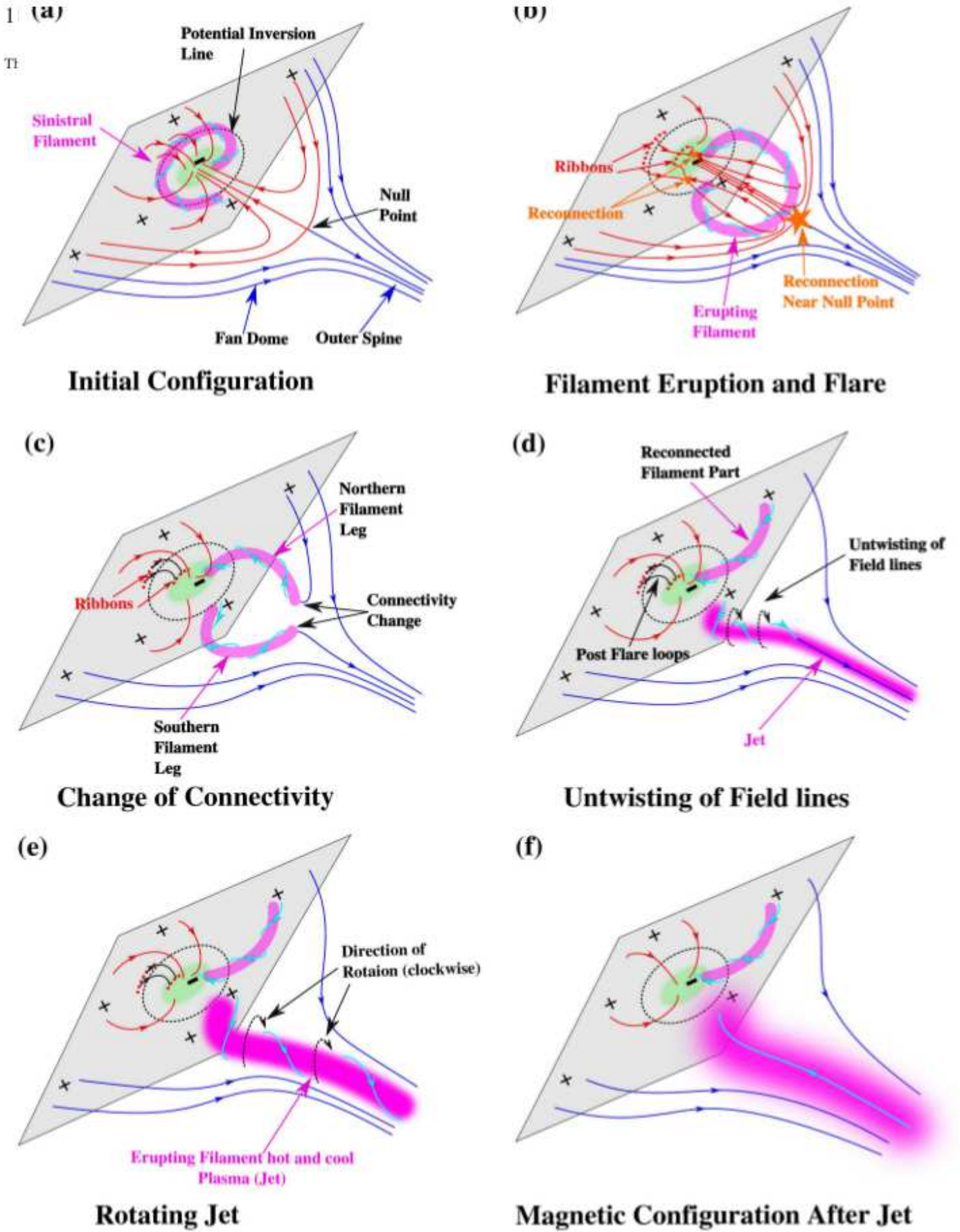}
	}
\vspace*{-3cm}
\caption{Schematic representation showing the different stages of the blowout Jet formation.}
\label{fig12}
\end{figure}

We analyzed a helical blowout jet which occurred on 2016 January 9. We discussed the interaction of the erupting mini-filament with the surrounding fan-spine structure, the origin of the rotation motion, and the associated CME in a context of the helical jet formation. We interpret the formation of the blowout jet as shown in the scheme in Figure~\ref{fig12}. A mini-filament (pink color in Figure~\ref{fig12}) is located inside the fan dome of a fan-spine configuration (c.f., Figures~\ref{fig1}(a)--~\ref{fig1}(d),~\ref{fig2},~\ref{fig3}, and~\ref{fig12}(a)). According to flux rope models \citep[][and references cited therein]{Martin98,Martin03,Filippov15}, the sinistral mini-filament is located inside a flux rope with the right-handed helix (cyan color in Figure~\ref{fig12}(a)). 

The blowout jet formation can be subdivided into three processes. First, the mini-filament is ejected upward by the instability and is accompanied by internal reconnection inside the fan dome. Secondly, the erupting filament interacts with the surrounding fan-spine structure (Figure~\ref{fig12}(c)). At that time, the eruptive filament changes from the arch shape to the cusp shape and forms a radial jet (Figures~\ref{fig5}(a)--(c)). Thirdly, the rotation motion is generated as the radial jet is developing. The untwisting motion of the twisted filament with the right--handed helix produced the clockwise rotation (Figures~\ref{fig7}, and~\ref{fig12}(e)). This untwisting motion straightens the field lines and at the end, only open magnetic field configuration is left (Figures~\ref{fig5}(f), and~\ref{fig12}(f)).

The present event is one of the clear examples which show overall steps of the formation and trigger processes of a blowout jet with high spatial and temporal resolutions. The event was initiated by a mini-filament eruption, triggered by flux convergence/cancellation in the quiet region, and followed by a helical jet formation. The presently studied jet is consistent with the new model for coronal jets that was first deduced by \cite{Adams14} for a macrospicule in a coronal hole and was first deduced by \cite{Sterling15} for all or nearly all blowout jets in coronal holes. Recently, \cite{Panesar16} found that the same model fits blowout jets in non-coronal-hole regions such as quiet regions, which are similar to the jet occurring region in our study.

\cite{Shibata86} discussed an unwinding motion of a twisted flux rope emerging into the open coronal fields as an origin of an H$\alpha$ surge (see also \citep{Kurokawa87}). In their scenario, one leg of a flux rope gets combined with the ambient magnetic field to form a closed loop, while the other leg gets connected with the open field line to form a jet with a helical motion \citep[see also][]{Filippov15a}. This is consistent with our observation (Figure~\ref{fig12}). It should be noted here that \cite{Shibata86} and \citep{Kurokawa87} did not consider the fan-spine configuration as the ambient field. On the other hand, \cite{Filippov15a} discussed the scenario in a fan-spine configuration, as in our observation and interpretation. 

The rotation motions of blowout jets have been observed and interpreted as the untwisting of core flux ropes \citep[e.g.,][]{Li15,Moore15}. The transverse velocity of a helical blowout jet is reported like $\sim$95 $km~s^{-1}$ by \cite{Hong11}, 25-70 $km~s^{-1}$ by \cite{Liu11}, 60-220 $km~s^{-1}$ by \cite{Moore15}, and 100-180 $km~s^{-1}$ by \cite{Filippov15a}. These are comparable to our analysis ($\sim$100 $km~s^{-1}$) and macrospicule observations \citep{Pike97,Parenti02,Pike98,Kamio10,Scullion09,Canfield96,Jiang07,Patsourakos08,Shen11,Lee13}.

More interestingly, we observed that the jet evolved like a standard jet in the initial phase from 13:43 UT to 13:51 UT. The twist manifests itself at 13:52 UT when the twisted southern leg starts unwinding. The 9-minutes time lag between the start times of the jet and the unwinding can be explained by the difference of the acceleration time scales. The jet is initially accelerated by magnetic reconnection in the corona and redirected by the ambient open magnetic fields. The time scale is determined by the local Alfven time scale, i.e. $\tau = L/v_{A} = (10^{10}~cm)/(1000~km~s^{-1}) = 100~s$. On the other hand, the unwinding motion is controlled by the smaller magnetic tension force and decelerated by dense cool plasma inside the filament. This makes the acceleration time scale of the unwinding motion much larger than the one of the jet.

We observed a narrow CME associated with the jet in the LASCO C2 and C3 fields of view (Figure~\ref{fig11}) that reached up to $\sim9~R_{sun}$. The speed of the CME is observed to be around 630 $km~s^{-1}$ which is faster than the initial jet speed ($\sim$394 $km~s^{-1}$). The transfer of magnetic helicity (or twist) from the low solar atmosphere to the corona pushes the jet material outward with increasing speed. The plasma acceleration mechanism (sweeping magnetic twist) in the jet is given by \cite{Shibata86}. They suggested that the untwisting motion produces the centrifugal force that accelerates the jet plasma. Later, \cite{Kurokawa87} confirmed the scenario of a sweeping magnetic twist by analysing an erupting untwisting filament.

The generation of the helical motion of a jet is related to the generation of waves along magnetic field lines. This is an important issue for coronal heating and solar wind acceleration. The helical motion is commonly observed in macrospicules and also in spicules, and we revealed the generation process of the helical motion in the extreme case. The physical process of a helical blowout jet is related to the one of macrospicules \citep{Kamio10,Kayshap13,Adams14}, which will be a key to reveal the helical motion generation of spicules and more fundamental physics of coronal streamers.

The exact process of reconnection between the erupting flux rope and the overlying fan--spine structure to trigger a blowout jets needs to be advanced. Also, the source of the rotation in blowout jets and the transport of helicity have to be discussed in detail. There is no simulation model available for this type of helical blowout jets with filament eruptions (and associated flux ropes) in fan--spine configurations. High--resolution observational study and numerical simulations are essential to understanding the overall process of formation and triggering of helical blowout jets in this kind of 3D magnetic configurations.


\section{Conclusions}

The main findings of this work are as follows.

\begin{enumerate}

\item We calculated the fan-spine configuration above the monopole magnetic field area and found a mini-filament underlying the fan-spine configuration before a helical blowout jet eruption.
\item The mini-filament erupted upward due to the magnetic flux-rope instability followed by internal reconnection, forming a two-ribbon structure at the footpoints of the post-flare loops. The filament interacted with the fan-spine magnetic structure, producing a helical blowout jet. This is consistent with an initiation of a blowout jet in models deduced by \cite{Adams14} and \cite{Sterling15}.
\item The jet shows a radial fast flow (~400 $km~s^{-1}$) and a horizontal rotation motion (~60-120 $km~s^{-1}$). The rotation motion of the jet is driven by the unwinding of the twist of the erupting mini-filament. The helicity of the sinistral filament matched the helicity of the rotating jet. 
\item The high speed of the associated narrow CME is the consequences of the magnetic reconnection and the helicity transfer from the lower to higher corona.


\end{enumerate}

\section*{Acknowledgements}

We thank the referee for his/her valuable comments and suggestions. We acknowledge the use of data taken by AIA and HMI on board SDO and LASCO on board SOHO. This work is supported by the BK21 plus program through the National Research Foundation (NRF) funded by the Ministry of Education of Korea. NCJ thanks, School of Space Research, Kyung Hee University for providing the postdoctoral grant. We are thankful to Drs. Sterling and Moore for their discussions and suggestions.







\clearpage


\clearpage


\clearpage


\clearpage


\clearpage


\clearpage


\clearpage


\clearpage

\clearpage

\clearpage


\clearpage


\clearpage



\bsp	
\label{lastpage}
\end{document}